\documentclass[twocolumn,superscriptaddress,amsmath,amssymb,prc,aps]{revtex4-1}

\usepackage{graphicx}  
\usepackage{siunitx}  
\usepackage{epsfig} 
\usepackage{subfigure}
\usepackage{lineno}  
\usepackage{amsmath,amssymb}
\usepackage{comment}
\usepackage{multirow}
\usepackage{dcolumn}% Align table columns on decimal point
\usepackage{bm}% bold math
\usepackage{setspace}
\usepackage{multibib}
\includecomment{printrobustnotes}
\includecomment{printallnotes}
\usepackage{xspace}
\usepackage{color}
\usepackage{xcolor}
\newcommand{\gevcc}[1]  {\ensuremath{#1~\mathrm{GeV}/{\rm c}^{2}}}
\begin{document}

\title{Strong constraints from COSINE-100 on the DAMA dark matter results using the same sodium iodide target}

\author{G.~Adhikari}
\affiliation{Department of Physics, University of California San Diego, La Jolla, CA 92093, USA}
\author{E.~Barbosa~de~Souza}
\affiliation{Department of Physics and Wright Laboratory, Yale University, New Haven, CT 06520, USA}
\author{N.~Carlin}
\affiliation{Physics Institute, University of S\~{a}o Paulo, 05508-090, S\~{a}o Paulo, Brazil}
\author{J.~J.~Choi}
\affiliation{Department of Physics and Astronomy, Seoul National University, Seoul 08826, Republic of Korea} 
\author{S.~Choi}
\affiliation{Department of Physics and Astronomy, Seoul National University, Seoul 08826, Republic of Korea} 
\author{M.~Djamal}
\affiliation{Department of Physics, Bandung Institute of Technology, Bandung 40132, Indonesia}
\author{A.~C.~Ezeribe}
\affiliation{Department of Physics and Astronomy, University of Sheffield, Sheffield S3 7RH, United Kingdom}
\author{L.~E.~Fran{\c c}a}
\affiliation{Physics Institute, University of S\~{a}o Paulo, 05508-090, S\~{a}o Paulo, Brazil}
\author{C.~Ha}
\affiliation{Department of Physics, Chung-Ang University, Seoul 06973, Republic of Korea}
\author{I.~S.~Hahn}
\affiliation{Department of Science Education, Ewha Womans University, Seoul 03760, Republic of Korea} 
\affiliation{Center for Exotic Nuclear Studies, Institute for Basic Science (IBS), Daejeon 34126, Republic of Korea}
\affiliation{IBS School, University of Science and Technology (UST), Daejeon 34113, Republic of Korea}
\author{E.~J.~Jeon}
\affiliation{Center for Underground Physics, Institute for Basic Science (IBS), Daejeon 34126, Republic of Korea}
\author{J.~H.~Jo}
\affiliation{Department of Physics and Wright Laboratory, Yale University, New Haven, CT 06520, USA}
\author{H.~W.~Joo}
\affiliation{Department of Physics and Astronomy, Seoul National University, Seoul 08826, Republic of Korea} 
\author{W.~G.~Kang}
\affiliation{Center for Underground Physics, Institute for Basic Science (IBS), Daejeon 34126, Republic of Korea}
\author{M.~Kauer}
\affiliation{Department of Physics and Wisconsin IceCube Particle Astrophysics Center, University of Wisconsin-Madison, Madison, WI 53706, USA}
\author{H.~Kim}
\affiliation{Center for Underground Physics, Institute for Basic Science (IBS), Daejeon 34126, Republic of Korea}
\author{H.~J.~Kim}
\affiliation{Department of Physics, Kyungpook National University, Daegu 41566, Republic of Korea}
\author{K.~W.~Kim}
\affiliation{Center for Underground Physics, Institute for Basic Science (IBS), Daejeon 34126, Republic of Korea}
\author{S.~H.~Kim}
\affiliation{Center for Underground Physics, Institute for Basic Science (IBS), Daejeon 34126, Republic of Korea}
\author{S.~K.~Kim}
\affiliation{Department of Physics and Astronomy, Seoul National University, Seoul 08826, Republic of Korea}
\author{W.~K.~Kim}
\affiliation{IBS School, University of Science and Technology (UST), Daejeon 34113, Republic of Korea}
\affiliation{Center for Underground Physics, Institute for Basic Science (IBS), Daejeon 34126, Republic of Korea}
\author{Y.~D.~Kim}
\affiliation{Center for Underground Physics, Institute for Basic Science (IBS), Daejeon 34126, Republic of Korea}
\affiliation{Department of Physics, Sejong University, Seoul 05006, Republic of Korea}
\affiliation{IBS School, University of Science and Technology (UST), Daejeon 34113, Republic of Korea}
\author{Y.~H.~Kim}
\affiliation{Center for Underground Physics, Institute for Basic Science (IBS), Daejeon 34126, Republic of Korea}
\affiliation{Korea Research Institute of Standards and Science, Daejeon 34113, Republic of Korea}
\affiliation{IBS School, University of Science and Technology (UST), Daejeon 34113, Republic of Korea}
\author{Y.~J.~Ko}
\email{yjko@ibs.re.kr}
\affiliation{Center for Underground Physics, Institute for Basic Science (IBS), Daejeon 34126, Republic of Korea}
\author{E.~K.~Lee}
\affiliation{Center for Underground Physics, Institute for Basic Science (IBS), Daejeon 34126, Republic of Korea}
\author{H.~Lee}
\affiliation{IBS School, University of Science and Technology (UST), Daejeon 34113, Republic of Korea}
\affiliation{Center for Underground Physics, Institute for Basic Science (IBS), Daejeon 34126, Republic of Korea}
\author{H.~S.~Lee}
\email{hyunsulee@ibs.re.kr}
\affiliation{Center for Underground Physics, Institute for Basic Science (IBS), Daejeon 34126, Republic of Korea}
\affiliation{IBS School, University of Science and Technology (UST), Daejeon 34113, Republic of Korea}
\author{H.~Y.~Lee}
\affiliation{Center for Underground Physics, Institute for Basic Science (IBS), Daejeon 34126, Republic of Korea}
\author{I.~S.~Lee}
\affiliation{Center for Underground Physics, Institute for Basic Science (IBS), Daejeon 34126, Republic of Korea}
\author{J.~Lee}
\affiliation{Center for Underground Physics, Institute for Basic Science (IBS), Daejeon 34126, Republic of Korea}
\author{J.~Y.~Lee}
\affiliation{Department of Physics, Kyungpook National University, Daegu 41566, Republic of Korea}
\author{M.~H.~Lee}
\affiliation{Center for Underground Physics, Institute for Basic Science (IBS), Daejeon 34126, Republic of Korea}
\affiliation{IBS School, University of Science and Technology (UST), Daejeon 34113, Republic of Korea}
\author{S.~H.~Lee}
\affiliation{IBS School, University of Science and Technology (UST), Daejeon 34113, Republic of Korea}
\affiliation{Center for Underground Physics, Institute for Basic Science (IBS), Daejeon 34126, Republic of Korea}
\author{S.~M.~Lee}
\affiliation{Department of Physics and Astronomy, Seoul National University, Seoul 08826, Republic of Korea} 
\author{D.~S.~Leonard}
\affiliation{Center for Underground Physics, Institute for Basic Science (IBS), Daejeon 34126, Republic of Korea}
\author{B.~B.~Manzato}
\affiliation{Physics Institute, University of S\~{a}o Paulo, 05508-090, S\~{a}o Paulo, Brazil}
\author{R.~H.~Maruyama}
\affiliation{Department of Physics and Wright Laboratory, Yale University, New Haven, CT 06520, USA}
\author{R.~J.~Neal}
\affiliation{Department of Physics and Astronomy, University of Sheffield, Sheffield S3 7RH, United Kingdom}
\author{S.~L.~Olsen}
\affiliation{Center for Underground Physics, Institute for Basic Science (IBS), Daejeon 34126, Republic of Korea}
\author{B.~J.~Park}
\affiliation{IBS School, University of Science and Technology (UST), Daejeon 34113, Republic of Korea}
\affiliation{Center for Underground Physics, Institute for Basic Science (IBS), Daejeon 34126, Republic of Korea}
\author{H.~K.~Park}
\affiliation{Department of Accelerator Science, Korea University, Sejong 30019, Republic of Korea}
\author{H.~S.~Park}
\affiliation{Korea Research Institute of Standards and Science, Daejeon 34113, Republic of Korea}
\author{K.~S.~Park}
\affiliation{Center for Underground Physics, Institute for Basic Science (IBS), Daejeon 34126, Republic of Korea}
\author{R.~L.~C.~Pitta}
\affiliation{Physics Institute, University of S\~{a}o Paulo, 05508-090, S\~{a}o Paulo, Brazil}
\author{H.~Prihtiadi}
\affiliation{Center for Underground Physics, Institute for Basic Science (IBS), Daejeon 34126, Republic of Korea}
\author{S.~J.~Ra}
\affiliation{Center for Underground Physics, Institute for Basic Science (IBS), Daejeon 34126, Republic of Korea}
\author{C.~Rott}
\affiliation{Department of Physics, Sungkyunkwan University, Suwon 16419, Republic of Korea}
\author{K.~A.~Shin}
\affiliation{Center for Underground Physics, Institute for Basic Science (IBS), Daejeon 34126, Republic of Korea}
\author{A.~Scarff}
\affiliation{Department of Physics and Astronomy, University of Sheffield, Sheffield S3 7RH, United Kingdom}
\author{N.~J.~C.~Spooner}
\affiliation{Department of Physics and Astronomy, University of Sheffield, Sheffield S3 7RH, United Kingdom}
\author{W.~G.~Thompson}
\affiliation{Department of Physics and Wright Laboratory, Yale University, New Haven, CT 06520, USA}
\author{L.~Yang}
\affiliation{Department of Physics, University of California San Diego, La Jolla, CA 92093, USA}
\author{G.~H.~Yu}
\affiliation{Department of Physics, Sungkyunkwan University, Suwon 16419, Republic of Korea}
\collaboration{COSINE-100 Collaboration}

\date{\today}

\begin{abstract}
We present new constraints on dark matter interactions using 1.7 years of COSINE-100 data. The COSINE-100 experiment, consisting of 106 kg of tallium-doped sodium iodide (NaI(Tl)) target material, is aimed at testing DAMA's claim of dark matter observation using the same NaI(Tl) detectors. Improved event selection requirements, a more precise understanding of the detector background and the use of a larger data set considerably enhances the COSINE-100 sensitivity for dark matter detection. No signal consistent with the dark matter interaction is identified, and rules out model-dependent dark matter interpretations of the DAMA signals in the specific context of standard halo model with the same NaI(Tl) target for various interaction hypotheses.
\end{abstract}
\maketitle

\section*{Introduction}
Astronomical observations continue to indicate that the Universe is made mostly of non-luminous, invisible dark matter~\cite{Clowe:2006eq, Ade:2015xua}. Several types of new fundamental particles have been proposed as candidates for the dark matter~\cite{Baer:2014eja} such as weakly interacting massive particles~(WIMPs)~\cite{PhysRevLett.39.165,Goodman:1984dc}, but no definitive signal has been seen despite concerted efforts by many collaborations~\cite{Battaglieri:2017aum}. One exception is the much-debated claim by the DAMA collaboration of a statistically significant annual modulation in the event rate of their experiment~\cite{BERNABEI1998195,Bernabei:2013xsa, Bernabei:2018yyw} with a period and phase consistent with that expected from WIMP dark matter~\cite{Savage:2008er,Baum:2018ekm, Kang:2018qvz}. This is controversial because if it is interpreted as a signature for WIMP interactions, it conflicts with other direct search experiments~\cite{Aprile:2017yea,Aprile:2018dbl,Akerib:2018zoq} that report null signals in the regions of parameter space that are allowed by the DAMA signals. 

Several groups have been working to develop new experiments with the aim of reproducing or refuting DAMA's results using the same NaI(Tl) detector material~\cite{Kim:2014toa,Adhikari:2017esn,Suerfu:2019snq,Fushimi:2021mez,Amare:2021yyu}. The COSINE-100 experiment is one of them that is currently operating with 106\,kg of low-background sodium iodide crystals at the Yangyang underground laboratory~\cite{Adhikari:2018ljm,Adhikari:2019off}.	 An analysis of the initial 59.5\,days of COSINE-100 data showed that the annual modulation signal reported by DAMA is inconsistent with the explanation using spin-independent interaction between WIMPs and  sodium or iodine nuclei in the context of the standard halo model~\cite{Adhikari:2018ljm}. However, this first result left open interpretations using certain alternative dark matter models~\cite{Kang:2019fvz,Zhitnitsky:2019tbh}, dark matter halo distributions~\cite{Bernabei:2020mon}, and detector responses~\cite{Bernabei:2020mon,Ko:2019enb} that could allow room for consistency  between DAMA and COSINE-100. 
Model independent searches of an annual modulation signal using 1.7\,years data were also reported but were still not sensitive enough to conclusively challenge the DAMA observation~\cite{Adhikari:2019off}.  
Here we present results from an analysis of 1.7\,years of COSINE-100 data with improved event selection requirements and an energy threshold that has been reduced from 2\,keVee to 1\,keVee, where keVee is kiloelectron volt electron-equivalent energy~\cite{Adhikari:2020xxj}. We find an order of magnitude improvement in sensitivity, sufficient for the first time to strongly constrain these alternative scenarios, as well as to further strengthen the previously observed inconsistency with the WIMP-nucleon spin-independent interaction hypothesis~\cite{Adhikari:2018ljm}. 

\section*{Results}
\subsection*{Experiment}
COSINE-100 is located at the Yangyang Underground Laboratory in South Korea with a 700\,m rock overburden~\cite{Adhikari:2018ljm,Adhikari:2019off}. The experiment consists of eight low-background thallium-doped sodium iodide (NaI(Tl)) crystals arranged in a 4$\times$2 array with a total target mass of 106\,kg. The array is  immersed in 2,200\,L of liquid scintillator used to identify events induced by radioactive background sources that are inside or outside the crystals~\cite{Adhikari:2020asl}.  The liquid scintillator is surrounded by copper and lead shields, and plastic scintillators to reduce the background contribution from external radiation as well as tag and reject events associated with cosmic-ray muons~\cite{Prihtiadi:2017inr}. Each NaI(Tl) crystal is optically coupled to two photomultiplier tubes (PMTs) with the signals recorded as 8\,$\mu$s waveforms. A trigger is generated when a signal corresponding to one or more photoelectrons occurs in each PMT within a 200\,ns time window~\cite{Adhikari:2018fpo}.

\begin{figure}[!htb]
  \begin{center}
    \includegraphics[width=1.0\columnwidth]{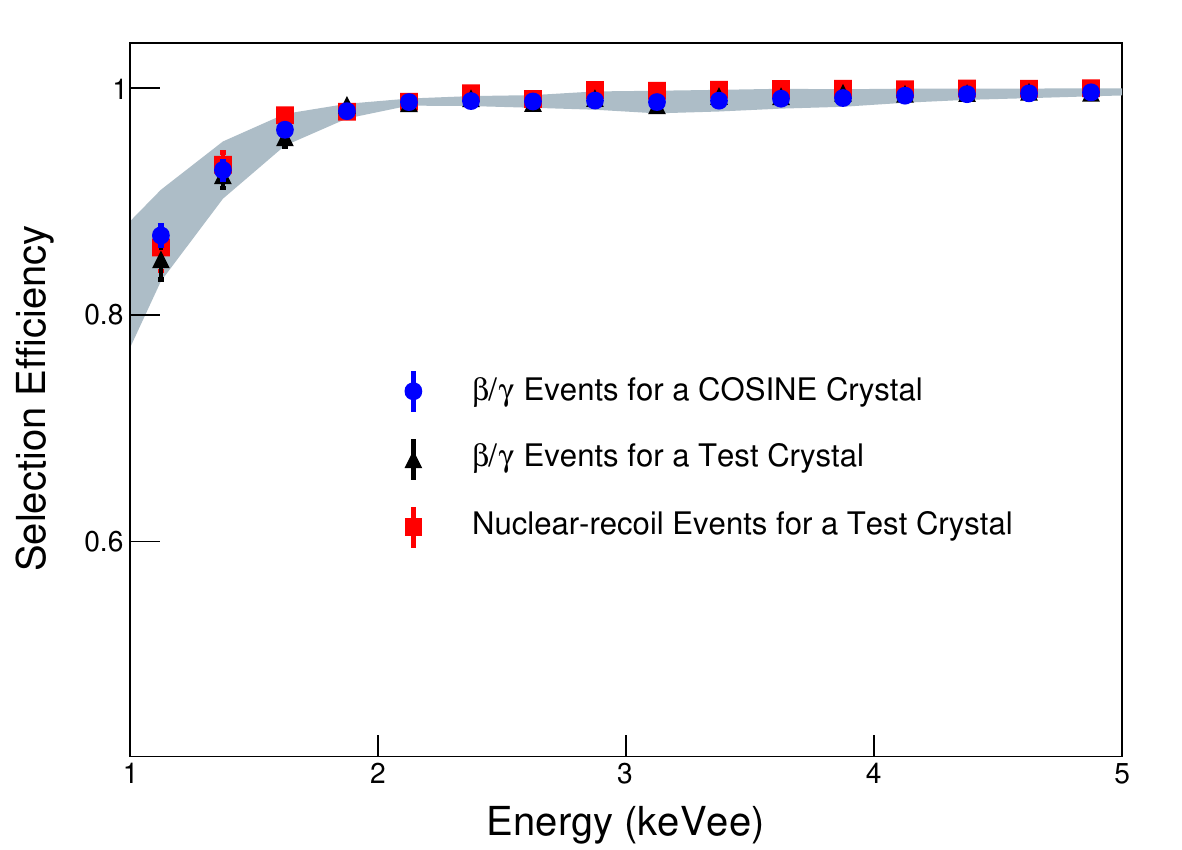} 
  \end{center}
  \caption{ 
    {\bf Efficiencies for $\beta$/$\gamma$ and nuclear-recoil events.}
    Blue dots show the efficiencies for $\beta$/$\gamma$ events for one of the COSINE-100 crystal. Black and red dots are efficiencies of $\beta$/$\gamma$ and nuclear-recoil events, respectively, for a small-size test crystals. This test crystal was cut from the same ingot of the COSINE-100 crystal and used for the neutron beam measurement. All measurements are consistent within the systematic uncertainty of the efficiency shown in grey band. 
}
  \label{fig:efficiency}
\end{figure}

The analysis presented here utilizes 1.7 years of data, previously used for the first annual modulation search~\cite{Adhikari:2019off}, and background modeling with a 1\,keVee energy threshold~\cite{cosinebg2}. The data were acquired between October 21, 2016 and July 18, 2018. Three of the eight crystals were observed to have high noise rates in the region of interest (ROI) and were excluded from the analysis, resulting in an effective data exposure of 97.7 kg$\cdot$year~\cite{Adhikari:2018ljm,Adhikari:2019off}.

It was found that PMT noise causes the majority of the triggered events in the ROI. A multivariable boosted decision tree (BDT)~\cite{BDT} was used to characterize the pulse-shapes to discriminate these PMT-induced noise events from radiation-induced scintillation events~\cite{Adhikari:2018ljm,Adhikari:2019off}. To improve the discrimination power, a likelihood score was introduced as an input training variable to the BDT that rates how well the waveform matches either scintillation events or PMT-induced noise events. The likelihood score particularly enhances the removal of noise pulses and allows us to comfortably operate with a 1 keVee threshold~\cite{Adhikari:2020xxj}. The BDT is trained with samples of scintillation-rich $^{60}$Co calibration data and PMT-noise dominant single-hit physics data. The multiple-hit events consist of in-time hits in multiple crystals or liquid scintillator that cannot be caused by dark matter interactions. The event selection efficiencies for scintillation events are evaluated with the $^{60}$Co calibration dataset and cross-checked with the physics data, as well as nuclear recoil events. The efficiencies from the $^{60}$Co calibration data were found to be consistent with previously measured efficiencies for nuclear recoil events obtained using a monoenergetic 2.42\,MeV neutron beam~\cite{Joo:2018hom} as shown in Fig.~\ref{fig:efficiency}. The efficiency differences and their uncertainties are included as a systematic uncertainty.

\subsection*{Background modeling}

\begin{figure*}[!htb]
  \begin{center}
    \includegraphics[width=1.0\textwidth]{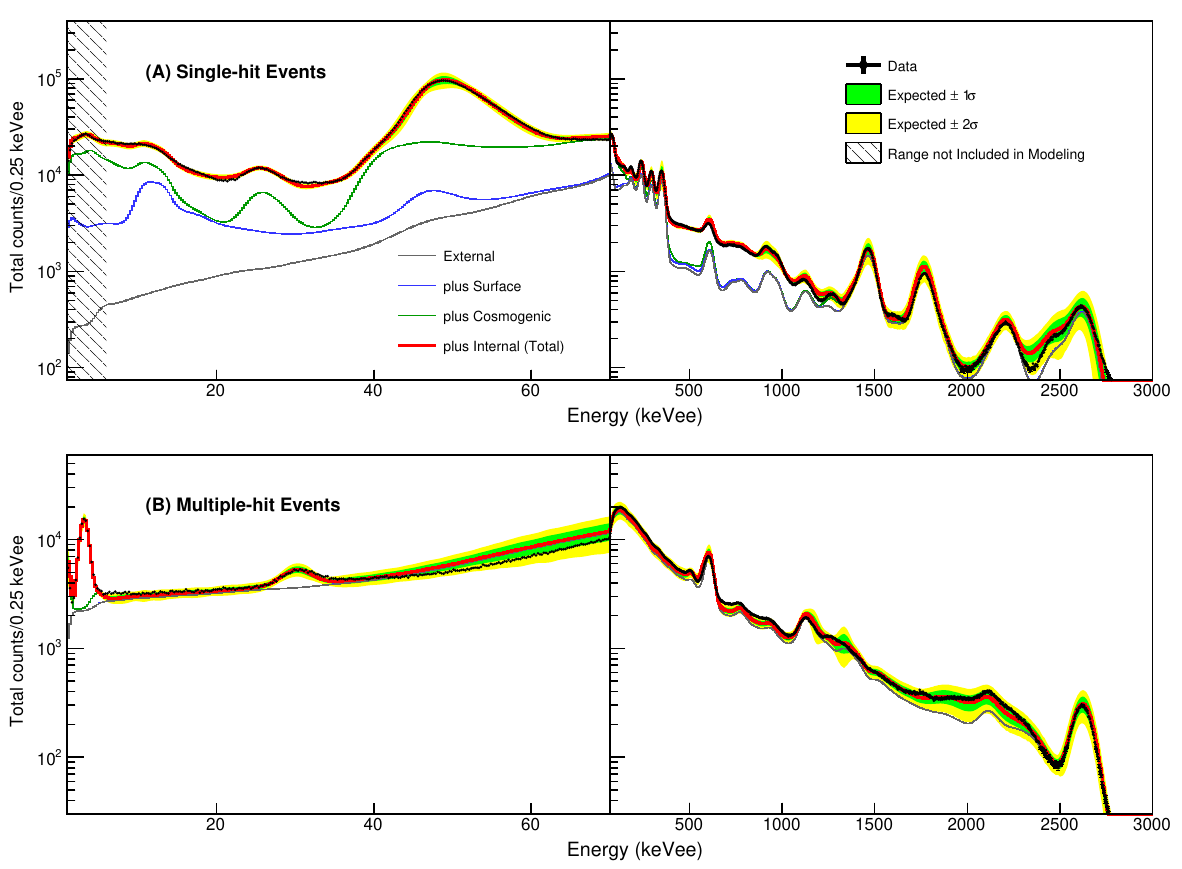} 
  \end{center}
  \caption{ 
    {\bf Energy spectra of single-hit and multiple-hit events.}
    Presented here are summed energy spectra for the five crystals (black dots) and their background models (red solid line) with the 68\% and 95\% confidence intervals. The expected contributions to the background from internal radionuclide contaminations, the surface of the crystals and nearby materials, cosmogenic activation, and external backgrounds are indicated. The 1--6\,keV region of the single-hit spectrum is masked because these events are not used for the background modeling.}
  \label{fig:background}
\end{figure*}

Events in the remaining dark matter search dataset predominantly originate from environmental $\gamma$ and $\beta$ radiations. Sources include radioactive contaminants internal to the crystals or on their surfaces, external detector components, and cosmogenic activation~\cite{cosinebg2}. In order to understand these events, the background spectrum for each individual crystal is modeled using computer simulations based on the Geant4 toolkit~\cite{Agostinelli:2002hh}.

Events are classified according to their energy: 1--70\,keVee are low energy and 70--3000\,keVee are high energy. The single-hit and multiple-hit data are separated in the background modeling of the NaI(Tl) crystals. To understand the background spectra, Geant4-based simulation events are generated and recorded in a format that matches that of the COSINE-100 data acquisition system. Energy resolutions and selection efficiencies for each crystal are applied. The fraction of each background component is determined from a simultaneous fit to the four measured distributions. For the single-hit events, only 6--3000\,keVee events are used to avoid a bias of the WIMP signal in the ROI. Details of the background modeling for the dataset are described elsewhere~\cite{cosinebg2}. 

The background components are divided into four categories: internal contamination, surface contamination, external sources and cosmogenic activation. The $^{238}$U, $^{232}$Th, $^{40}$K, and $^{210}$Pb contaminations in the crystal constitute the internal background.  The $^{210}$Pb contaminations on the crystal surface and adjacent materials are the surface component. Backgrounds from $^{238}$U, $^{232}$Th and $^{40}$K in the PMTs, liquid scintillator, and the shield materials constitute the external sources. In order to estimate contributions from cosmogenic activation, we use a time-dependent analysis that takes into account the cosmic-ray exposure time on the ground and the cooling time in the underground laboratory of each individual crystal~\cite{deSouza:2019hpk}.

The most dominant background components in the ROI are generated by internal radionuclide contamination and by cosmogenic activation. This includes $^{210}$Pb and $^{40}$K internal contaminants, and $^{210}$Pb surface contamination. The contribution to the ROI from cosmogenic activation is mostly due to $^3$H with some additional contributions from $^{113}$Sn and $^{109}$Cd. Background modeling was performed independently for each individual crystal, and Fig.~\ref{fig:background} shows the accumulated result of the model fit to data and the systematic uncertainties.

Several sources of systematic uncertainty are identified and included in this analysis. The largest systematic uncertainties are those associated with the efficiencies, which include statistical errors in the efficiency determinations with the  $^{60}$Co calibration and systematic errors derived from the independent cross-checks of the physics data and the nuclear recoil events. Uncertainties in the energy resolution and nonlinear responses of the NaI(Tl) crystals~\cite{nonprop} affect the shapes of the background and signal spectra. The depth-profiles of $^{210}$Pb on the surface of the NaI(Tl) crystals, studied with a $^{222}$Rn contaminated crystal, are varied within their uncertainty~\cite{Yu:2020ntl}. Variations in the levels and the positions of external Uranium and Thorium decay-chain contaminants are also taken into account. Effects of event rate variations and possible distortions in the shapes of spectra are considered in systematic uncertainties. 

\subsection*{Dark matter interpretations}
We consider various WIMP models to determine the possible contribution from WIMP interactions to the measured energy spectra using the simulated data. The DAMA/LIBRA-phase2 data~\cite{Bernabei:2018yyw} were found not to be compatible with the canonical model~\cite{Baum:2018ekm,Ko:2019enb}, which is an isospin-conserving spin-independent interaction between WIMP and nucleus in the specific context of the standard WIMP galactic halo model, and is the most commonly used interpretation of the direct detection of the WIMP dark matter~\cite{Tanabashi:2018oca}. However, an isospin-violating interaction in which the WIMP-proton coupling is different from the WIMP-neutron coupling, provides a good fit to the observed annual modulation signals from the DAMA/LIBRA-phase2 data~\cite{Baum:2018ekm,Ko:2019enb}. To interpret the DAMA/LIBRA data and compare with the COSINE-100 data, we use the best fit values  of the effective coupling of WIMPs to neutrons and to protons~($f_n/f_p$) obtained for the simultaneous fit of DAMA/LIBRA-phase1 and DAMA/LIBRA-phase2 data described elsewhere~\cite{Ko:2019enb}. We also interpret the results of the COSINE-100 data in the canonical model for the comparison with the DAMA/LIBRA-phase1 only data. 

We use the nuclear recoil quenching factor (QF) from recent measurements with monoenergetic neutron beams~\cite{Joo:2018hom} (quenching factor is the ratio of the scintillation light yield from sodium or iodine recoil relative to that for electron recoil for the same energy). In those measurements, neutron tagging detectors at a fixed angle relative to the incoming neutron beam direction provide unambiguous knowledge of the deposited energy. We obtained a strong energy dependence of the nuclear recoil QFs. 
Modelings of the QF measurements described in Ref.~\cite{Ko:2019enb} are appropriated for this analysis (subsequently referred to as new QF). 
However,  most studies interpreting the DAMA/LIBRA's results have used significantly larger QF values that were reported by the DAMA group in 1996~\cite{BERNABEI1996757} (subsequently referred to as DAMA QF), that were obtained by measuring the response of NaI(Tl) crystals to nuclear recoils induced by neutrons from a $^{252}$Cf source. 
The best description of the measured nuclear recoil spectra from the $^{252}$Cf source obtained 30$\pm$1\% and 9$\pm$1\% for the sodium and iodine QF values, respectively, assuming no energy dependence of the QF values. The values obtained from the new measurements are approximately 13\% and 5\% for the sodium and iodine, repsectively, at 20\,keVnr where keVnr is kiloelectron volt nuclear recoil energy~\cite{Ko:2019enb}. 
%Assuming no energy dependence of the nuclear recoil QFs and. 
%For example, the sodium and iodine QF values were reported by DAMA to be 30\% and 9\%, while the new measurements find these to be approximately 13\% and 5\%, respectively, at 20\,keVnr where keVnr is nuclear recoil energy~\cite{Ko:2019enb}. 
Efficient noise rejection as well as correct evaluation of trigger and selection efficiencies are essential for proper estimation of the quenching factors~\cite{Collar:2013gu,Xu:2015wha,Joo:2018hom}. 
%The measurements of the DAMA QF values were required to check the trigger and selection efficiencies in low-energy regions and consider energy dependent QF as pointed in Ref.~\cite{Collar:2013gu}.
Even though the measurements of the DAMA QF values were required to check the efficiency evaluations as well as no energy-dependent QF assumption~\cite{Collar:2013gu}, 
the hypothesis of different QFs~\cite{Bernabei:2020mon} in the NaI(Tl) crystals used by DAMA/LIBRA and COSINE-100 needs to be checked. 
Note that results from the analysis of the previous 59.5\,days of COSINE-100 data with a 2\,keVee threshold were not sufficient to exclude all the DAMA/LIBRA 3$\sigma$ regions when different QFs are used~\cite{Ko:2019enb}.

\begin{figure}[!htb]
  \begin{center}
    \includegraphics[width=1.0\columnwidth]{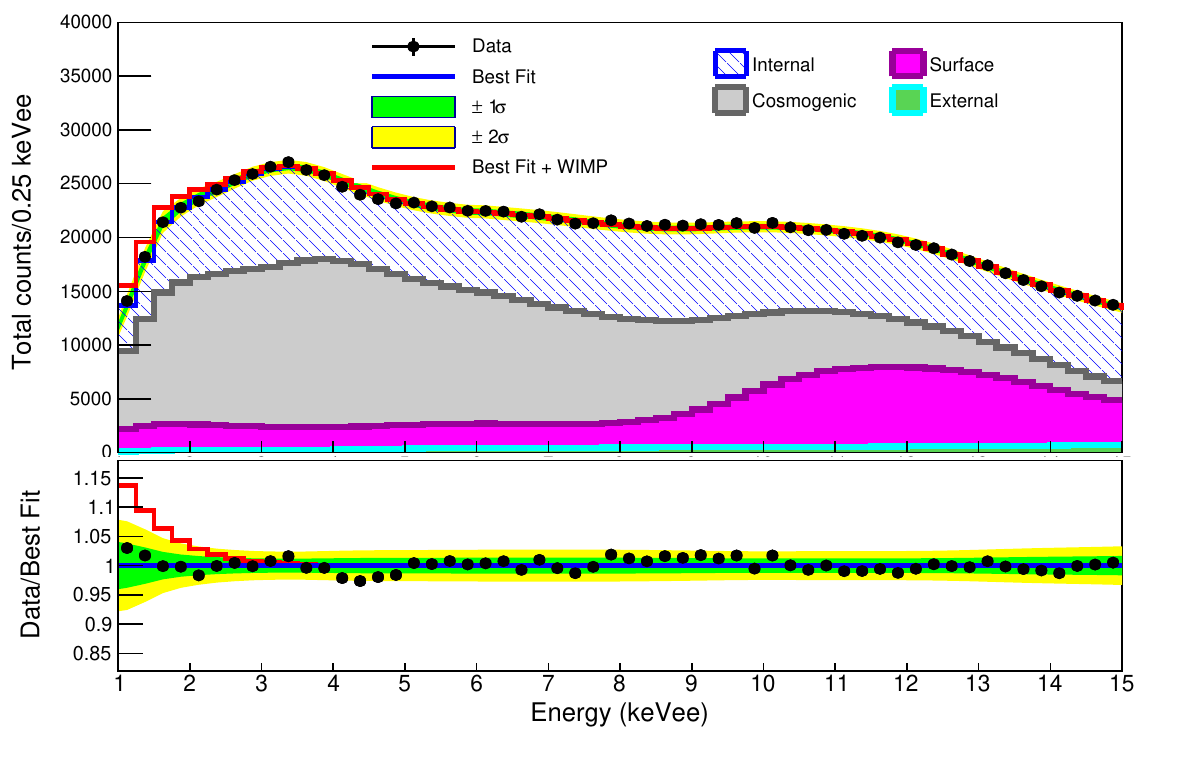} 
  \end{center}
  \caption{ 
    {\bf Example fit results for a 11.5\,GeV/c$^2$ WIMP mass in the case of $\bm{f_n/f_p = -0.76}$ .}
    Presented here is the summed energy spectrum for the five crystals (black filled circles shown with 68\% confidence level error bars) and the best fit (blue line) for which no WIMP signals are obtained.  Fitted contributions to the background from internal radionuclide contaminations, the surface of the crystals and nearby materials, cosmogenic activation, and external backgrounds are indicated. The green (yellow) bands are the 68\% (95\%) confidence level intervals of the systematic uncertainty obtained from the likelihood fit. For presentation purposes, we indicate the signal shape (red line) assuming a WIMP-proton cross section of 2.5$\times 10^{-2}$\,pb corresponding to the DAMA best fit value for the WIMP-sodium interaction using the DAMA QF values.}
  \label{fig:fit}
\end{figure}

\begin{figure*}[!htb]
  \begin{center}
    \includegraphics[width=1.0\textwidth]{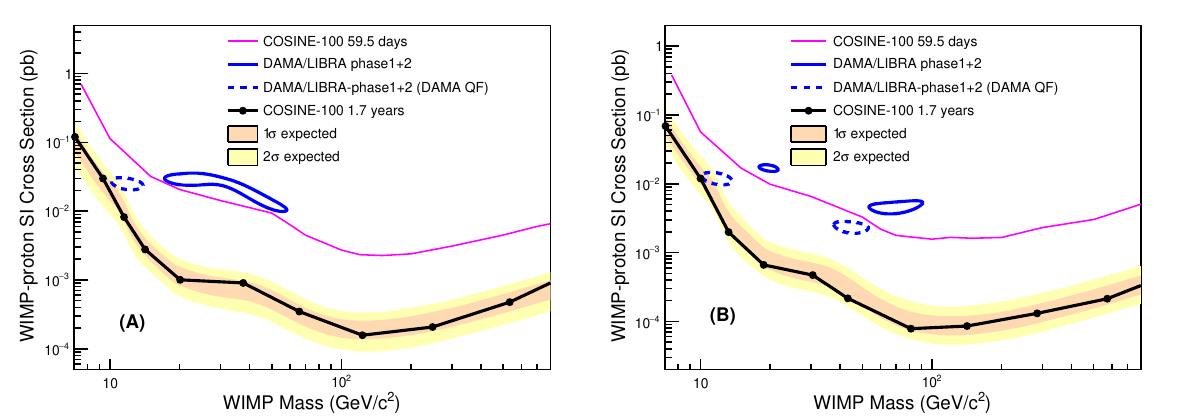}
  \end{center}
  \caption{
    {\bf Exclusion limits on the WIMP-proton spin-independent cross section for the isospin-violating interaction.}
    The 3$\sigma$ allowed regions of the WIMP mass and the WIMP-proton cross-section associated with the DAMA/LIBRA-phase1+phase2 data (blue solid coutours) using the new QF values in their best fit for (A) sodium scattering and (B) iodine scattering hypotheses are compared with the 90\% confidence level exclusion limits from the COSINE-100 data (black-solid-line), together with the 68\% and 95\% probability bands for the expected 90\% confidence level limit assuming the background-only hypothesis. The dashed blue contours show the allowed regions of the DAMA/LIBRA-phase1+phase2 data using the DAMA QF values. For comparison, limits from the initial 59.5\,days COSINE-100 data~\cite{Adhikari:2018ljm} are shown by the purple-solid-line. In each plot, we fix the effective coupling ratios to neutrons and protons $f_n/f_p$ to the best fit values of the DAMA data.}
\label{results_iv}
\end{figure*}

To search for evidence of a WIMP signal in the data, a Bayesian approach with a likelihood function based on Poisson probability is used. The likelihood fit is applied to the measured single-hit energy spectra between 1 and 15\,keVee for each WIMP model for several masses. Each crystal is fitted with a crystal-specific background model and a crystal-correlated WIMP signal for the combined fit by  multiplying the five crystals' likelihoods. Means and uncertainties for background components, which are determined from the modeling~\cite{cosinebg2}, are used to set Gaussian priors for the background. The systematic uncertainties are included in the fit as nuisance parameters with Gaussian priors (see the section of materials and methods).

A good fit to the DAMA/LIBRA-phase2 data was obtained with the isospin-violating interaction~\cite{Baum:2018ekm,Ko:2019enb}. We simultaneously use the DAMA/LIBRA-phase1 and phase2 data to fit three parameters: the WIMP mass, the WIMP-proton cross-section, and $f_n/f_p$. The best fits were obtained for two different values of $f_n/f_p$ favoring WIMP-sodium and WIMP-iodine interactions as $f_n/f_p = -0.76$ and $-0.71$, respectively. For the best fit values of $f_n/f_p$, the 3$\sigma$ allowed regions in the WIMP-mass and the WIMP-proton cross-section parameter spaces are obtained~\cite{Ko:2019enb}.

The COSINE-100 data are fitted to each of the different WIMP masses for each $f_n/f_p$ value using only the new QF values. An example of a maximum likelihood fit for a \gevcc{11.5} WIMP and $f_n/f_p=-0.76$  WIMP signal is presented in Fig.~\ref{fig:fit}. The summed event spectrum for the five crystals is shown together with the best-fit result. For comparison, the expected signal for a \gevcc{11.5} WIMP  with a spin-independent WIMP-proton cross section of $2.5\times10^{-2}$\,pb, the central value of the DAMA/LIBRA best fit using the DAMA QF values for the WIMP-sodium interaction, is shown by the red solid line. No excess of events that could be attributed to WIMP interactions is found for the considered WIMP signals. The posterior probabilities of signals are consistent with zero in all cases and 90\% confidence level limits are determined (see Fig.~S\ref{fig:posterior}). Figure~\ref{results_iv} shows the 3$\sigma$ contours of the DAMA/LIBRA data in the best fit values of $f_n/f_p$ using the new QF values and the DAMA QF values together with the 90\% confidence level upper limits from the COSINE-100 data using the same $f_n/f_p$ and the new QF values. The 90\% confidence level limits from the 1.7 years COSINE-100 data show approximately an order of magnitude better limits than those of our previous results using 59.5\,days data and exclude the DAMA/LIBRA allowed 3$\sigma$ regions for both sets of QF values.

\begin{figure}[!htb]
  \begin{center}
    \includegraphics[width=1.0\columnwidth]{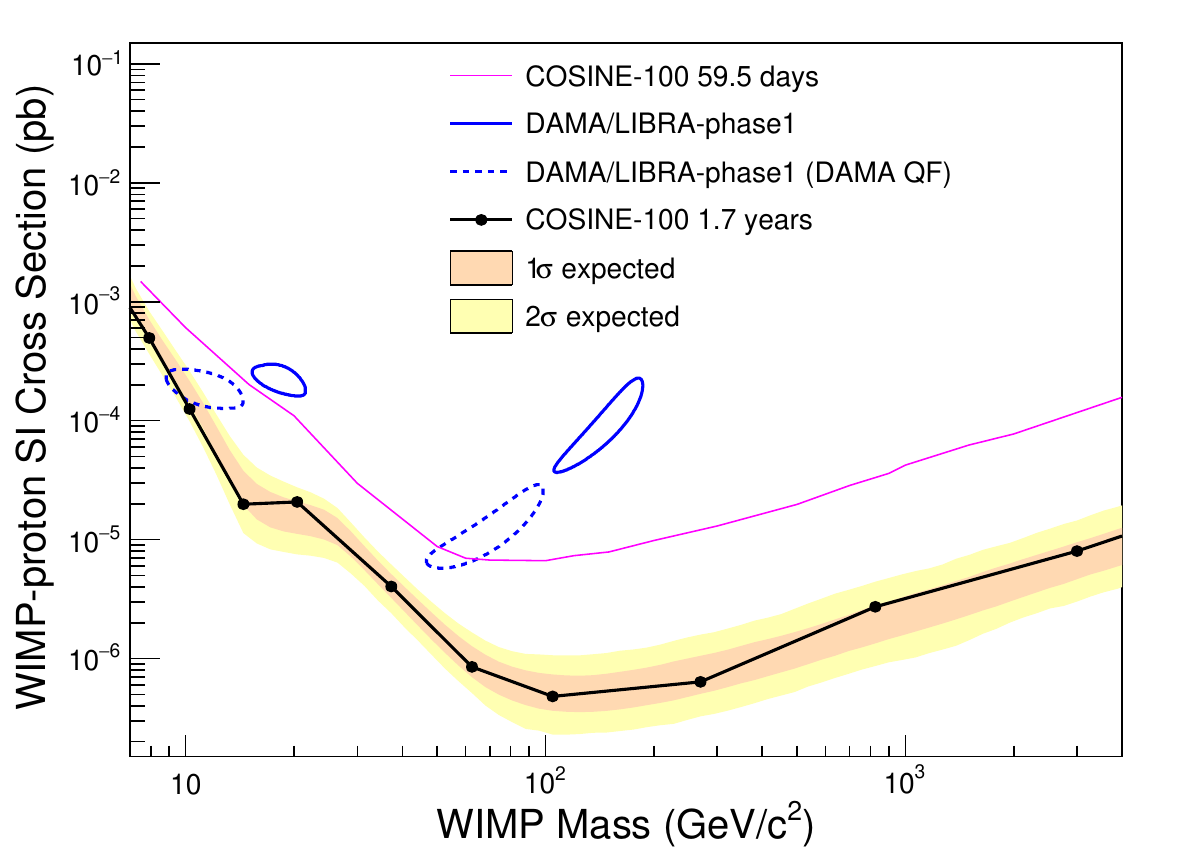}
  \end{center}
  \caption{
    {\bf Exclusion limits on the WIMP-nucleon spin-independent cross section of the isospin-conserving interaction.}
    The observed (filled circles with black solid line) 90\% confidence level exclusion limits on the WIMP-nucleon spin-independent cross section from the COSINE-100 are shown together with the 68\% and 95\% probability bands for the expected 90\% confidence level limit, assuming the background-only hypothesis.  The limits are compared with a WIMP interpretation of the DAMA/LIBRA-phase1 3$\sigma$ allowed region using the new QF (blue-solid-contours) and the DAMA QF (blue-dashed-contours)~\cite{Savage:2008er}.
}
\label{results_ic}
\end{figure}

Even though the DAMA/LIBRA-phase2 data do not fit well to the canonical model, their phase1 data has been shown to be well fit with an isospin-conserving spin-independent WIMP-nuclei interaction~\cite{Savage:2008er,Ko:2019enb}.  The 90\% confidence level upper limits from the COSINE-100 data for the canonical model are also obtained. Figure~\ref{results_ic} shows the 3$\sigma$ allowed regions that are associated with the DAMA/LIBRA-phase1 signal using the new QF values and the DAMA QF values together with the 90\% confidence level upper limits from the COSINE-100 data using the new QF values. These limits mostly exclude the DAMA/LIBRA allowed region even when different QF values are considered for each experiment. 

\begin{figure*}[!htb]
  \begin{center}
    \includegraphics[width=0.8\textwidth]{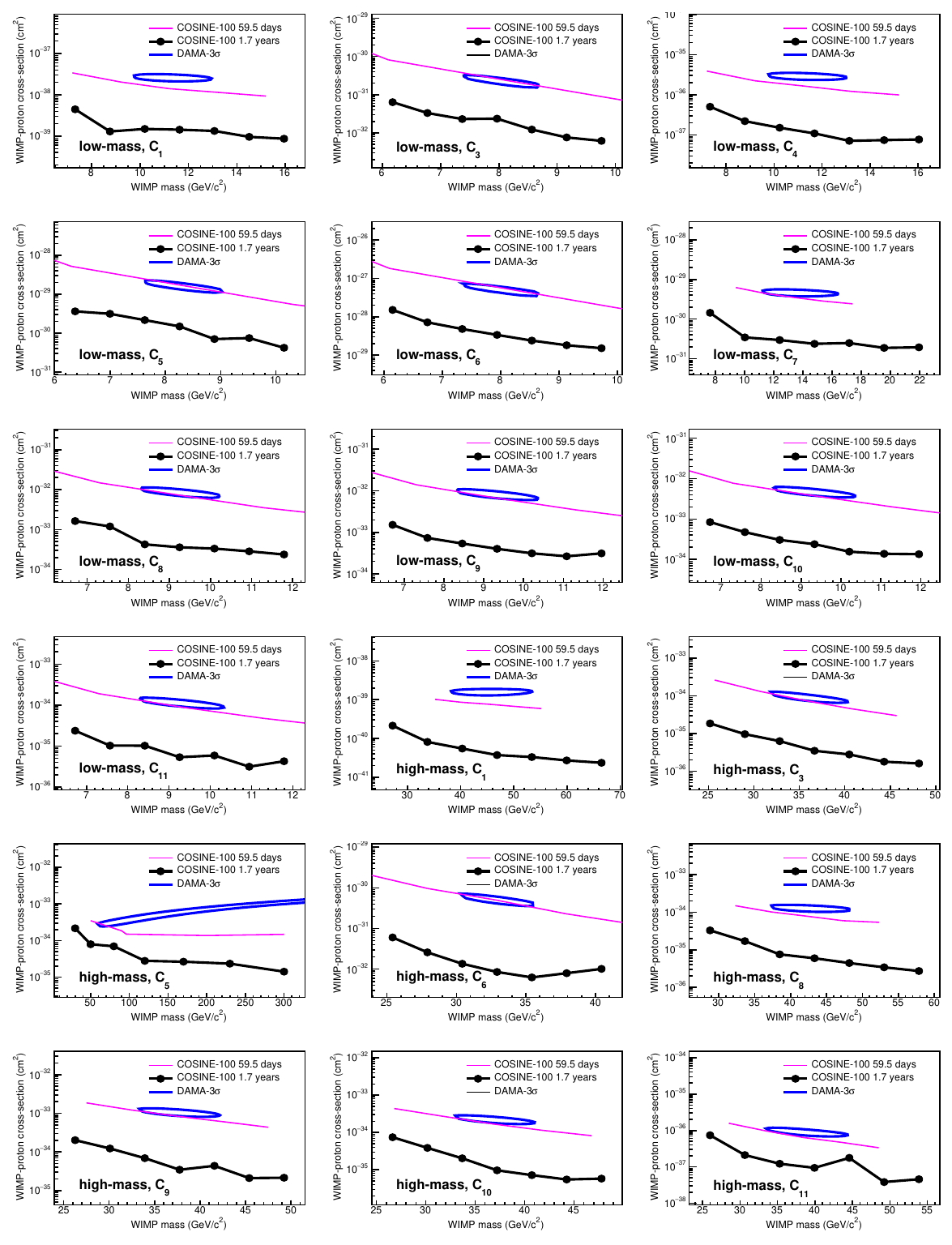} 
  \end{center}
  \caption{ 
    {\bf Exclusion limits on the WIMP-proton cross section for the effective field theory operators.}
    DAMA/LIBRA 3$\sigma$ allowed regions (blue contours) and COSINE-100 90\% confidence level exclusion limits of previous analysis (pink solid lines) and this work (black dots and lines)	on the WIMP-proton cross sections for a variety of effective field theory operators using the DAMA QF values are presented.	For each operator, $f_n/f_p$ is fixed to the corresponding best fit value of the DAMA/LIBRA data.}
  \label{fig:eft}
\end{figure*}

In addition, we have checked each operator in an assortmenet of non-relativistic effective field theory models where previous null results from the 59.5\,days COSINE-100 data do not fully cover the 3\,$\sigma$ regions of the DAMA/LIBRA data for a few operators~\cite{Kang:2019fvz}. The 1.7\,years data is now found to fully cover the 3\,$\sigma$ allowed regions for each model assuming the DAMA QF values,  as can be seen in Fig.~\ref{fig:eft}. 

\section*{Discussion}

After the release of the initial 59.5\,days COSINE-100 data with null observations using the same NaI(Tl) target material, a few possibilities were suggested that preserve the consistency between the  DAMA/LIBRA and COSINE-100 results~\cite{Kang:2019fvz,Ko:2019enb,Bernabei:2020mon}. The results of this analysis, with 1.7\,years accumulated COSINE-100 data and improved analysis technique with a 1\,keVee energy threshold do not favor these suggested possibilities. A model independent data analysis of the annual modulation with several years COSINE-100 data is required for an unambiguous conclusion, nevertheless the results presented here provide strong constraints on the dark matter interpretation of the DAMA/LIBRA annual modulation signals with the same NaI(Tl) target materials.

\section*{Acknowledgments}
We thank the Korea Hydro and Nuclear Power (KHNP) Company for providing underground laboratory space at Yangyang. This work is supported by:  the Institute for Basic Science (IBS) under project code IBS-R016-A1 and NRF-2021R1A2C3010989, Republic of Korea; NSF Grants No. PHY-1913742, DGE-1122492, WIPAC, the Wisconsin Alumni Research Foundation, United States; STFC Grant ST/N000277/1 and ST/K001337/1, United Kingdom; Grant No. 2017/02952-0 FAPESP, CAPES Finance Code 001, CNPq 131152/2020-3, Brazil.

\clearpage

\section*{Appendix}

\subsection*{event selection}

An event satisfying the trigger condition of coincident photoelectrons in both of the crystal's readout PMTs within 200\,ns is acquired with 500\,MHz flash analog-to-digital converters and recorded as a 8\,$\mu$s long waveform starting 2.4\,$\mu$s before occurrence of the trigger~\cite{Adhikari:2018fpo}. In the offline analysis, muon-induced events are rejected when the crystal hit event occurs within 30\,ms after a muon candidate event in the muon detector~\cite{Prihtiadi:2017inr,Prihtiadi:2020yhz}. Additionally, we require that leading edges of the trigger pulses start later than 2.0\,$\mu$s after the start of the recording, waveforms from the crystal contain more than two single photoelectrons, and the integral waveform area below the baseline does not exceed a limit. These criteria reject muon-induced phosphor events and electronic interference events. A multiple-hit event has accompanying crystal signals with more than four photoelectrons in an 8\,$\mu$s time window or has a liquid scintillator signal above an 80\,keVee threshold within 4\,$\mu$s~\cite{Adhikari:2020asl}. A single-hit event is classified as one where the other detectors do not meet these criteria.

During the 1.7\,years data-taking period, no significant environmental abnormality or unstable detector performance was observed. The high light yield of the six crystals, approximately 15 photoelectrons/keVee, allowed an analysis threshold of 2\,keVee in the previous analysis. However, the other two crystals had lower light yields and required higher analysis thresholds of 4\,keVee and 8\,keVee respectively~\cite{Adhikari:2017esn,Adhikari:2018ljm}. Since their direct impact on the WIMP search is not substantial, we do not include single-hit events from these two crystals in the WIMP search analysis, although they were used in the identification of multiple hits.

In the low-energy signal region below 10\,keVee, PMT-induced noise events predominantly contribute to the single-hit physics data in two different ways. The first class has a fast decay time of less than 50\,ns compared with typical NaI(Tl) scintillation of about 250\,ns. The second class, that occurs less often than the first, has different characteristics of slow rise and decay time, as characterized in Refs.~\cite{Adhikari:2018ljm,Kang:2019fvz}. Noise events of the second class are intermittently produced by certain PMTs. We have developed monitoring tools for data quality verification, including monitoring event rates of the second class of noise. If a crystal had an increased rate due to the second class of noise, the relevant period of data was removed. One crystal detector had this class of noise for the whole period, but for the other five detectors more than 95\,\% of the recorded data could be used without the second-class noise-induced events. 

A boosted decision tree~(BDT) was developed to separate scintillation signals from the first class of noise. The fast PMT-induced events with energies greater than 2\,keVee were efficiently removed by the BDT, which is based on multiple parameters~\cite{Adhikari:2018ljm,Kang:2019fvz} the balance of the deposited charge from two PMTs, the ratio of the leading-edge (0--50\,ns) to trailing-edge (100\,ns--600\,ns) charge, and the amplitude weighted average time of the signal. However, with this BDT, the scintillation events with energies below 2\,keVee were contaminated by an exponential increase in noise events.

We developed new parameters to characterize the PMT-induced noise events for efficient selection of the scintillation events below 2\,keVee. Two likelihood parameters for an event are constructed by templates derived from data samples enriched alternatively in scintillation-signal events and noise-signal events. A $^{60}$Co source that produces low energy electron signals through Compton scattering is used to generate the signal enriched sample. Fast decay events in the single-hit data sample are used as the noise enriched sample.

\begin{figure}[!htb]
  \begin{center}
    \includegraphics[width=1.0\columnwidth]{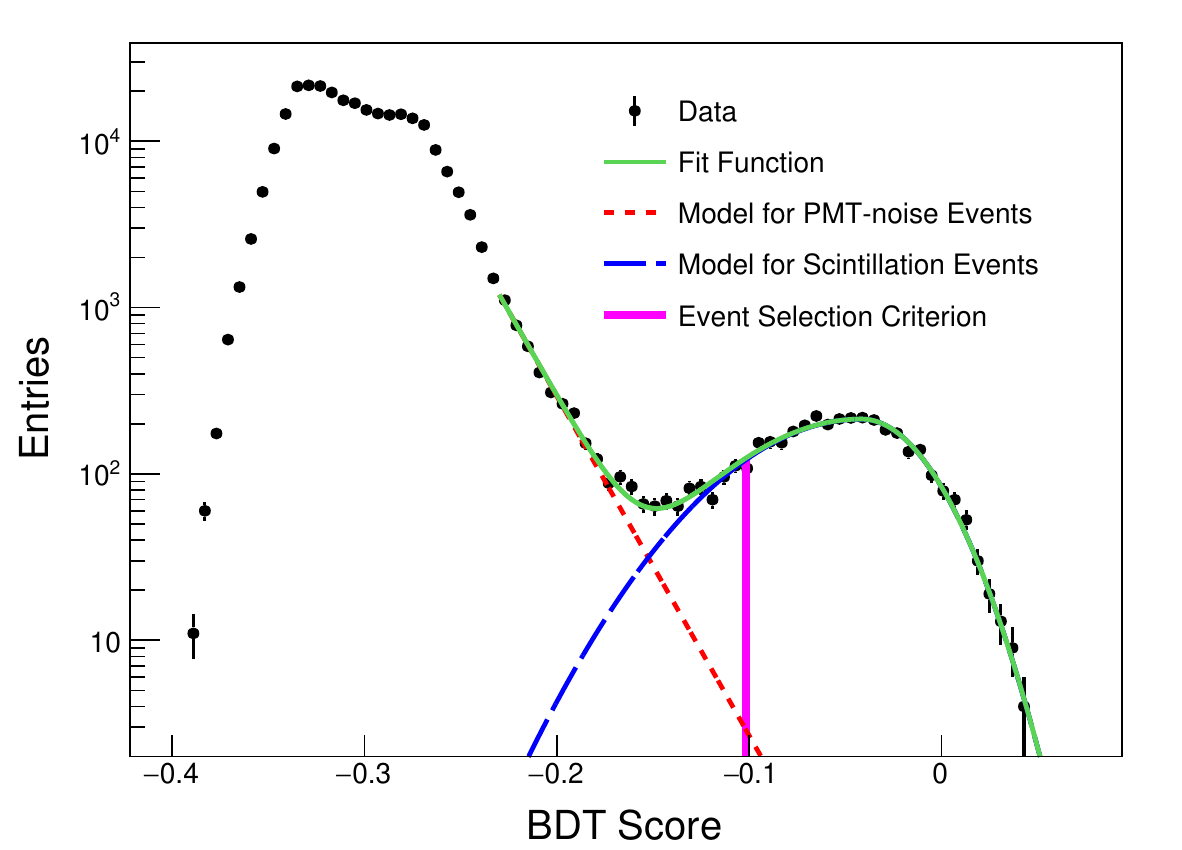} 
  \end{center}
  \caption{ 
    {\bf BDT-score distribution of events at the 1--1.25\,keV.}
    The BDT distribution of the single-hit physics data is fitted by the green function that consists of an asymmetric Gaussian distribution for the scintillation events (dashed blue) and an exponential function for the PMT-noise events (dotted red). The thick magenta line shows a BDT criterion for the event selection.}
  \label{fig:bdt_dist_cut}
\end{figure}

The likelihood parameter of the event is calculated as,
\begin{equation}
  \ln\mathcal{L} = \sum_i\left[T_i - W_i + W_i\ln\frac{W_i}{T_i}\right],
\end{equation}
where $T_i$ and $W_i$ are the height of the $i^{th}$ time bin in the reference template and event, respectively. Likelihood parameters for the scintillation signal events ($\ln\mathcal{L}_s$) and the PMT-induced noise events ($\ln\mathcal{L}_n$) are evaluated for each event. 
With these, we define a likelihood score as, 
\begin{eqnarray}
  p_l=\frac{\ln\mathcal{L}_n - \ln\mathcal{L}_s}{\ln\mathcal{L}_n + \ln\mathcal{L}_s},
\end{eqnarray}
where large $p_l$ for an event implies that the event is more closely matched to the scintillation signal than the PMT-induced noise events. The updated BDT is trained with the likelihood score. This provided good discrimination against PMT-induced noise events and enabled us to lower the threshold to 1\,keVee~\cite{Adhikari:2020xxj}. A BDT score, a single discriminating parameter created by combining the various selections for input parameters according to their corresponding importance, for the single-hit physics data near the energy threshold (1--1.25\,keVee) is presented in Fig.\ref{fig:bdt_dist_cut}. With these selection criteria, we reduced the PMT-induced noise contamination level to less than $0.5$\,\%.

Event selection efficiencies for the electron recoil events were evaluated with the $^{60}$Co calibration data. The efficiency is calculated by integrating the model distribution for the scintillation signals and the PMT-induced noise events shown in Fig.~\ref{fig:bdt_dist_cut}.  A specialized apparatus with a monoenergetic 2.42\,MeV neutron beam was used to measure the selection efficiencies of the nuclear recoil events~\cite{Joo:2018hom}. This measurement was performed with a small-size test crystal that was cut from the same ingot as a crystal used for the COSINE-100 experiment. The efficiencies determined for the electron recoil events and the nuclear recoil events are consistent within the 5\% level as shown in Fig.~\ref{fig:efficiency}. Systematic uncertainties in the efficiency measurements account for deviations from different measurements, as well as from the $^{60}$Co calibration data and the single-hit physics data.

\subsection*{systematic uncertainties}

In addition to the statistical uncertainties in the background and signal models, various sources of systematic uncertainties are taken into account. Errors in the selection efficiency, the energy resolution, the energy scale, and background modeling technique translate into uncertainties in the shapes of the signal and background probability density functions, as well as to rate changes. These quantities are allowed to vary within their uncertainties as nuisance parameters in the likelihood fit. 

\begin{figure}[!htb]
  \begin{center}
    \includegraphics[width=1.0\columnwidth]{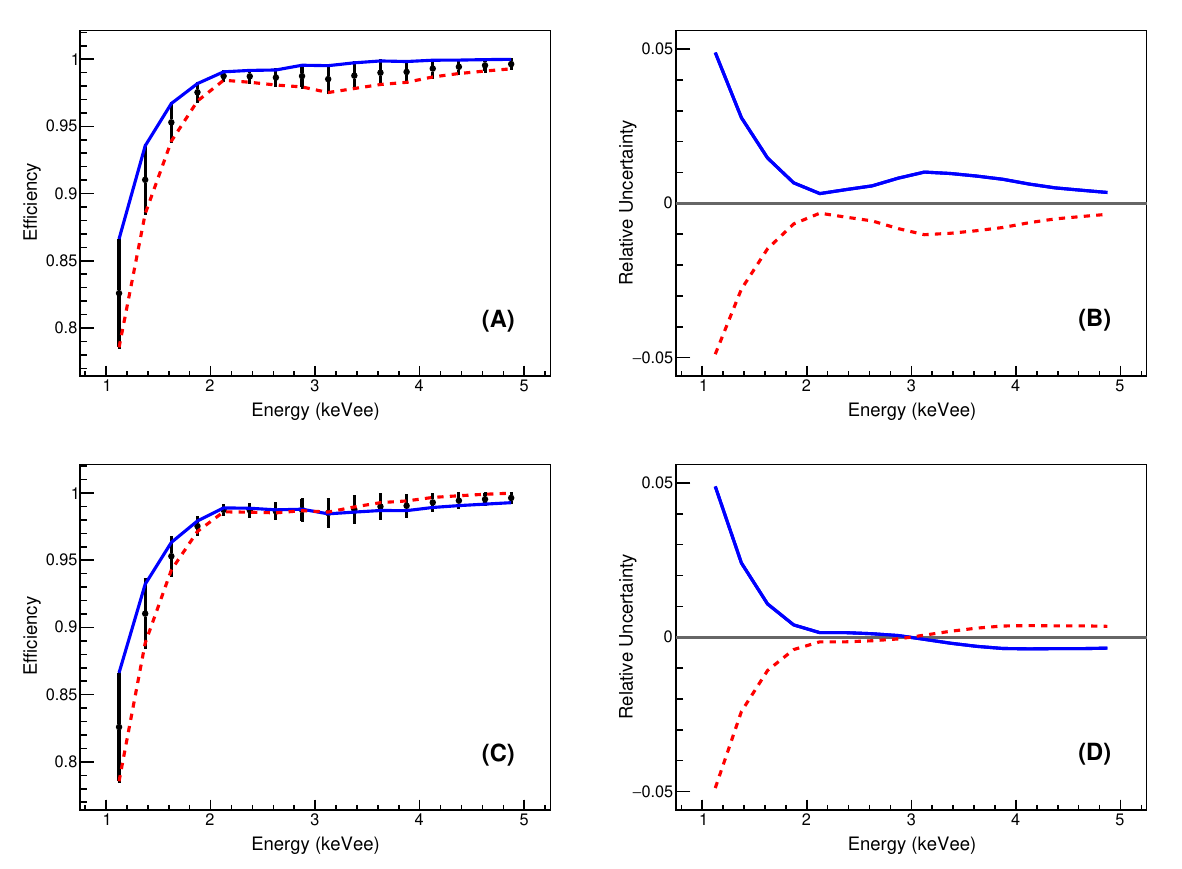} 
  \end{center}
  \caption{ 
    {\bf Systematic uncertainty of the event selection efficiency}
    Black dots with error bars in (A) and (C) present the event selection efficiency and associated systematic uncertainty.  (A) The upper (blue-solid line) and lower (red-dashed line) limits in $1\sigma$ uncertainties and (B) their associated relative uncertainty are shown. In addition, the maximum distortions of shape that can mimic the WIMP signal are accounted (C). Here the blue-solid line starts from the upper $1\sigma$ in the first energy bin (1--1.25\,keV) and evenly moves to the lower $-1\sigma$ in the last energy bin (4.75--5\,keV). The red-dashed line represent the opposite changes of the efficiency systematic. (D) Relative uncertainties associated with (C) are shown. The efficiency systematic is accounted as nuissance parameters in the likelihood using above two-types of the relative uncertainties.}
  \label{fig:eff_syst}
\end{figure}

The most influential systematic uncertainty is the error associated with the efficiencies shown as the shaded region in Fig.~\ref{fig:efficiency}. This is because the efficiency systematic uncertainty maximally covers the statistical uncertainties in the ROI. We include two types of systematic variations in the efficiency error. At first, we account maximum variations of the uncertainty across the width of the energy bin to provide $\pm 1\sigma$ ranges as shown in Fig.~\ref{fig:eff_syst} (A). Relative uncertainties shown in Fig.~\ref{fig:eff_syst} (B) are included as nuisance parameters of the background and signal in the likelihood function (see Eq.~\ref{eq:bkgd_nuisance} and Eq.~\ref{eq:sig_nuisance}). However, due to the dominant errors in the ROI, we consider maximum shape distortions that can mimic the WIMP signal as shown in Fig.~\ref{fig:eff_syst} (C). Its relative uncertainty is shown in Fig.~\ref{fig:eff_syst} (D) and included as a nuisance parameter.

In the background model fit, the levels of background activities are limited by Gaussian constraint terms added to the likelihood function as determined by measured activities and their uncertainties. The systematic uncertainties associated with the background modeling include the uncertainties of the activities estimated by the background model fit. In addition, different locations of external radioactive contaminations are taken into account by generating external contributions at different positions. Background contributions from $^{210}$Pb contamination on the surface of the NaI(Tl) crystals were studied with a small NaI(Tl) crystal exposed to $^{222}$Rn from a $^{226}$Ra source~\cite{Yu:2020ntl}. Depth profiles from two exponential components were modeled to fit the $^{222}$Rn contaminated crystal and matched to the test-setup data~\cite{cosinebg2}. Uncertainties in the measured depth profiles are propagated into systematic uncertainties. We generate the background model associated with each systematic variation and account relative uncertainty to be added as the nuissance parameter. 

The energy calibration is performed by tracking the positions of internal $\beta$ and $\gamma$ peaks from radioactive contaminations in the crystals, as well as with external $\gamma$ sources~\cite{cosinebg2}. The nonlinear detector response of the NaI(Tl) crystals~\cite{nonprop} in the low energy region is modeled with an empirical function across all crystals~\cite{cosinebg2}. Subtle differences for each crystal from the general nonlinearity model of the NaI(Tl) crystals are evaluated to consider the systematic uncertainty on the energy scale. The energy resolution for each crystal is evaluated during the calibration process. In particular, tagged 0.8\,keVee ($^{22}$Na) and 3.2\,keVee ($^{40}$K) X-ray lines in the multiple-hit data are used to determine the energy resolution for low-energy events. Statistical uncertainties associated with the number of tagged events are regarded as the resolution systematic. Figure~\ref{fig:background} shows a comparison of the background model to the data together with $\pm$1$\sigma$ and $\pm$2$\sigma$ bands of the systematic uncertainties that are evaluated from the quadrature sum of each systematic component. The relative errors are used for the nuisance parameter of the background. 

\subsection*{expected WIMP signal}

The differential nuclear recoil rate per unit target mass for elastic scattering between WIMPs of mass $m_\chi$ and target nuclei of mass $M$ is~\cite{Savage:2008er},
\begin{eqnarray}
  \frac{dR}{dE_\mathrm{nr}} = \frac{\rho_\chi}{2m_\chi\mu^2}~\sigma(M,~E_\mathrm{nr})
  \int_{v>v_\mathrm{min}}d^3v~f(\textbf{v},~t),
\label{eq:recoilrate}
\end{eqnarray}
where $\rho_\chi$ is the local dark matter density, $E_\mathrm{nr}$ is the nuclear recoil energy, $\sigma(M,E_\mathrm{nr})$ is the WIMP-nucleus cross section and $f(\textbf{v},t)$ is the time-dependent WIMP velocity distribution. The reduced mass $\mu$ is defined as $m_\chi M/(m_\chi+M)$ and the minimum WIMP velocity $v_\mathrm{min}$ is $\sqrt{ME_\mathrm{nr}/2\mu^2}$.

For the WIMP velocity distribution, we assume  the Standard Halo Model (SHM)~\cite{freese1987},
\begin{eqnarray}
  f(\textbf{v},t) = \begin{cases}
  	\frac{1}{N_\mathrm{esc}}~\left(\frac{3}{2\pi\sigma_v^2}\right)^{3/2}&e^{-3[\textbf{v}+\textbf{v}_\mathrm{E}(t)]^2/2\sigma_v^2},\\
		& \mbox{for }\left|\textbf{v}+\textbf{v}_\mathrm{E}(t)\right|<v_{esc}\\
	0,&\mbox{otherwise,}
  \end{cases}
\label{eq:vdist}
\end{eqnarray}
where $N_\mathrm{esc}$ is a normalization constant, $\textbf{v}_E$ is the Earth velocity relative to the WIMP dark matter and $\sigma_v$ is the velocity dispersion. The SHM parameterization is used with the local dark matter density $\rho_\chi = 0.3~\mathrm{GeV/cm}^3$, $v_E = 232~\mathrm{km/s}$, $\sqrt{2/3}~\sigma_v = 220~\mathrm{km/s}$ and the galactic escape velocity $v_\mathrm{esc} = 544~\mathrm{km/s}$~\cite{ref:par_SHM}.

\begin{figure*}[!htb]
  \begin{center}
    \includegraphics[width=1.0\textwidth]{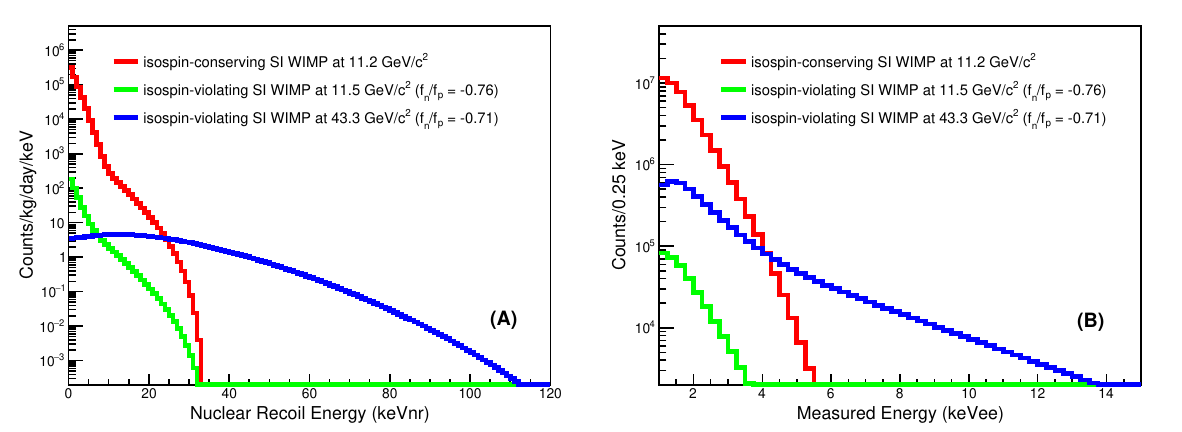} 
  \end{center}
  \caption{ 
    {\bf Energy spectra of the WIMP signal.}
		(A) Raw energy spectra for three different WIMP models with WIMP-proton cross section of 1\,pb. (B) The expected energy spectra with WIMP-proton cross section of 1\,pb for the three WIMP models when taking account of the quenching factors, detector resolutions, and selection efficiencies assuming 1.7 years COSINE-100 data.}
  \label{fig:signal}
\end{figure*}

The effective field theory operators and nuclear form factors described in Refs.~\cite{Gluscevic:2015sqa, Anand:2013yka, Fitzpatrick:2012ix, Gresham:2014vja} are used to model the nuclear responses in the differential cross section. The generalized spin-independent response~\cite{Fitzpatrick:2012ix} is used for both the isospin-conserving and the isospin-violating spin-independent (SI) interactions. For isospin-violating SI interactions, the WIMP-nucleon coupling coefficient ratio, $f_n/f_p$ is fixed to the best fit values for the DAMA/LIBRA data~\cite{Ko:2019enb}. These nuclear responses, including form factors, are implemented using the publicly available {\sc dmdd} package~\cite{dmdd} to evaluate the WIMP-nucleus cross section, $\sigma(M,E_\mathrm{nr})$ in Eq.~\ref{eq:recoilrate} (raw signal spectra). We subsequently apply the quenching factors, energy resolution, and selection efficiency to obtain the expected nuclear recoil rate in electron-equivalent energy for the detector, $dR/dE_\mathrm{ee}$. Figure~\ref{fig:signal} shows $dR/dE_\mathrm{nr}$ (A) and $dR/dE_\mathrm{ee}$ (B) spectra for three different WIMP models  assuming WIMP-proton cross section equal to 1\,pb. 

In addition, the WIMP signals are generated in the context of a non-relativistic effective theory of WIMP-nucleus scattering that has been tested using previous data without full coverage of the DAMA/LIBRA 3$\sigma$ allowed regions~\cite{Kang:2019fvz}. For simplicity, we assume that one of the effective operators allowed by Galilean invariance dominates in the effective Hamiltonian of a spin-half dark matter particle at a time. We use the best-fit neutron-over-proton coupling ratio of the DAMA/LIBRA-phase2 data assuming the DAMA QF reported in Ref.~\cite{Kang:2018qvz}, for each operator. The  {\sc dmdd} package~\cite{dmdd} is also used to generate signal spectra by the effective operators. A few operators are not evaluated because the {\sc dmdd} package does not include form factors for these operators.  Here, we assume the DAMA QFs for the COSINE-100 data for a simple comparison.

\subsection*{Bayesian approach}
A Bayesian approach is adopted to extract the WIMP signal from the COSINE-100 data. For each WIMP interaction model, a posterior probability density in terms of the WIMP-proton cross section $\sigma_\mathrm{WIMP}$ is obtained from the Bayes' theorem using a marginalization of the likelihood function that includes the prior~\cite{Tanabashi:2018oca},
\begin{equation}
		\begin{split}
  \mathcal{P}(\sigma_\mathrm{WIMP}|\mathbf{M}) = & N\int d\bm{\alpha}\int d\bm{\beta}~\mathcal{L}(\mathbf{M}|\sigma_\mathrm{WIMP},~\bm{\alpha},~\bm{\beta})\\ 
	                                               & \times \pi(\sigma_\mathrm{WIMP}), 
  \label{eq:posteriorPDF}
		\end{split}
\end{equation}
where $\mathcal{P}(\sigma|\mathbf{M})$ is a PDF and $\mathcal{L}(\mathbf{M}|\sigma,\bm{\alpha},\bm{\beta})$ is the likelihood function. The prior $\pi(\sigma_\mathrm{WIMP})$ has a constant and zero values for positive and negative $\sigma_\mathrm{WIMP}$, respectively. The normalization constant $N$ makes the integral of the posterior PDF unity and $\mathbf{M}$ represents the measured data. The $\bm{\alpha}$ and the $\bm{\beta}$ denote the nuisance parameters to control the effect by systematic uncertainties. Because the measurements are independent and follow Poisson probabilities, the likelihood function is built as a product of Poisson probabilities,
\begin{equation}
		\begin{split}
  \mathcal{L}(\mathbf{M}|\sigma_\mathrm{WIMP},~\bm{\alpha},~\bm{\beta})&  =  \prod_i^{N_\mathrm{crystal}}\prod_j^{N_\mathrm{bin}}
  \frac{[E_{ij}(\sigma_\mathrm{WIMP},~\bm{\alpha},~\bm{\beta})]^{M_{ij}}}{M_{ij}!} \\ 
  & \times e^{-E_{ij}(\sigma_\mathrm{WIMP},~\bm{\alpha},~\bm{\beta})}\cdot\pi(\bm{\alpha},~\bm{\beta}),
  \label{eq:likelihood}
		\end{split}
\end{equation}
where $i$ and $j$ denote the crystal number and the energy bin, respectively, and $M_{ij}$ is the number of observed events for crystal $i$ in  $j^\mathrm{th}$ energy bin. The $\pi(\bm{\alpha},\bm{\beta})$ is a term for constraining nuisance parameters by systematic uncertainties.

The expected number of events, denoted as $E_{ij}(\sigma,\bm{\alpha},\bm{\beta})$, is obtained as the sum of the number of events in the signal and background,
\begin{equation}
  E_{ij}(\sigma_\mathrm{WIMP},~\bm{\alpha},~\bm{\beta}) = S_{ij}(\sigma_\mathrm{WIMP},\bm{\alpha}) + B_{ij}(\bm{\alpha},~\bm{\beta}),
  \label{eq:expectednumber}
\end{equation}
where the number of background events $B_{ij}(\bm{\alpha},\bm{\beta})$ and signal events $S_{ij}(\sigma,\bm{\alpha})$ are generated from the simulated experiments through the background modeling and the WIMP signal discussed above, with effects by systematic uncertainties. The systematic uncertainty affecting the background model is included as a function of the nuisance parameter $\bm{\alpha}$ and $\bm{\beta}$, as
\begin{equation}
  B_{ij}(\bm{\alpha},~\bm{\beta}) = \prod_k^{N_\mathrm{syst}}(1+\alpha_{ik}\epsilon_{ijk})\prod_l^{N_\mathrm{bkgd}}(1+\beta_{il})\cdot B_{ij}^\mathrm{MC},
  \label{eq:bkgd_nuisance}
\end{equation}
where $B_{ij}^\mathrm{MC}$ is the number of background events obtained from the model. The nuisance parameter $\alpha_{ik}$ controls the effect of the energy-dependent uncertainty, $\epsilon_{ijk}$, which is 1$\sigma$ relative error for $k^\mathrm{th}$ systematic uncertainty. Meanwhile, another nuisance parameter $\beta_{il}$ adjusts the activity for $l^\mathrm{th}$ background component. The corresponding impact on the WIMP signal is considered by means of the expression,
\begin{equation}
  S_{ij}(\sigma_\mathrm{WIMP},~\bm{\alpha})
  = \prod_k^{N_\mathrm{syst}}(1+\alpha_{ik}\epsilon_{ijk})\cdot T_i\cdot M_i\cdot R_j(\sigma_\mathrm{WIMP};m_\chi),
  \label{eq:sig_nuisance}
\end{equation}
where $M_i$ and $T_i$ denote the mass and data exposure for crystal $i$, and $R_j$ is the expected rate of WIMP-proton interaction through an integration of $dR/dE_\mathrm{ee}$ in the $j^{\mathrm{th}}$ energy bin. Each nuisance parameter is constrained with evaluated uncertainty assuming a Gaussian distribution,
\begin{equation}
  \pi(\bm{\alpha},~\bm{\beta}) = \prod_i^{N_\mathrm{crystal}}\prod_k^{N_\mathrm{syst}}\exp\left[-\frac{\alpha_{ik}^2}{2}\right]\prod_l^{N_\mathrm{bkgd}}\exp\left[-\frac{\beta_{il}^2}{2\delta_{il}^2}\right],
  \label{eq:prior_syst}
\end{equation}
where $\delta_{il}$ is the uncertainty of the activity of the $l^\mathrm{th}$ background component. A Markov Chain Monte Carlo~\cite{ref:mcmc1,ref:mcmc2} via Metropolis-Hastings algorithm~\cite{ref:metropoils,ref:hastings} is used for the multivariable integration in posterior PDF. We developed our own Bayesian tool for this process. A comparison with a publicly available Bayesian analysis toolkit~\cite{Caldwell:2008fw} was done for the initial 59.5\,days COSINE-100 data and both tools showed consistent results.

\begin{figure*}[!htb]
  \begin{center}
    \includegraphics[width=1.0\textwidth]{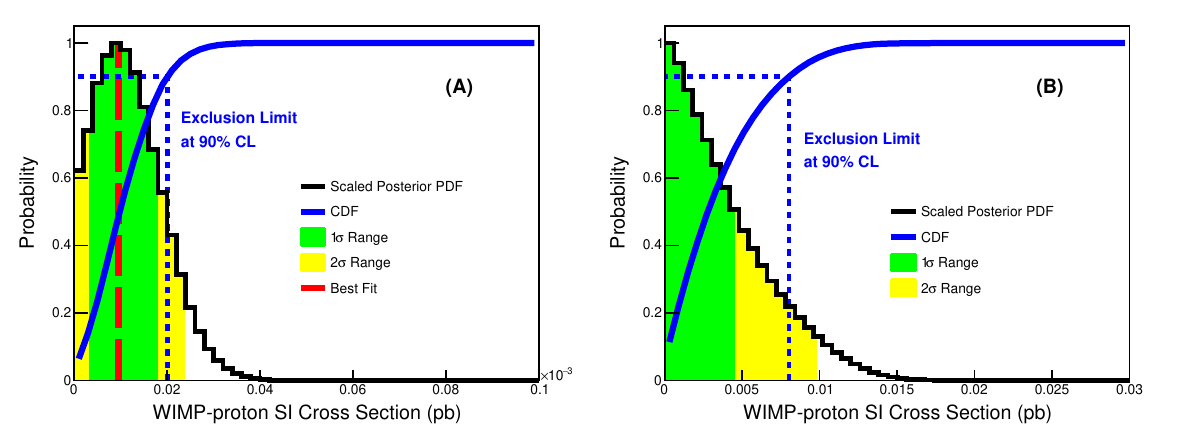} 
  \end{center}
  \caption{ 
    {\bf Posterior probability density functions.}
    Two examples of the posterior PDFs and their CDFs for the 1.7\,years COSINE-100 data for different WIMP models. (A) is the canonical model for a WIMP mass of 20.4~GeV/c$^2$ and (B) is the isospin-violating case with WIMP mass 11.5~GeV/c$^2$ and $f_n/f_p = -0.76$. The posterior PDFs are scaled for the maximum to be unity.  The exclusion limit at 90\% confidence level is obtained from CDF matched with 0.9. Green and yellow areas represent the 68\% and 95\% of confidence intervals, respectively. In the case of (A), the best fit value (red-dashed line) presents slightly positive result of WIMP-proton cross section 9.5$\times 10^{-5}$\,pb, but within 95\% region. We, therefore, set the 90\% confidence level upper limit.}
  \label{fig:posterior}
\end{figure*}

To avoid biasing the WIMP search, the fitter was tested with simulated event samples. Each experimental data is prepared by Poisson random extraction from the modeled background spectrum~\cite{cosinebg2}, assuming a background-only hypothesis. Marginalization to obtain the posterior PDF for each simulation sample is performed to set the 90\% confidence level exclusion limits as shown in Fig.~\ref{fig:posterior}. The 1000 simulated experiments result in 68\% and 95\% bands of the expected limit presented in Figs.~\ref{results_iv} and \ref{results_ic}. The data fits are done in the same way as the simulated data. Figure~\ref{fig:posterior} shows the posterior PDFs and their cumulative distribution functions (CDFs) of data for two different WIMP models. The CDF provides the 90\% confidence level exclusion limit for each fit.

\bibliographystyle{PRTitle} 
\providecommand{\href}[2]{#2}\begingroup\raggedright\endgroup

\end{document}